\documentclass[a4paper,11pt]{article}
\pdfoutput=1
\usepackage{jinstpub}
\usepackage{lineno}
\usepackage{tikz}
\usetikzlibrary{arrows.meta}
\usetikzlibrary{backgrounds}
\usetikzlibrary{shapes.geometric}
\usepackage{ifthen}
\usepackage{subcaption}
\usepackage[acronym]{glossaries}

\newacronym{ai}{AI}{Artificial Intelligence}
\newacronym{conv2d}{Conv2D}{2D Convolution}
\newacronym{cnn}{CNN}{Convolutional Neural Network}
\newacronym{dnn}{DNN}{Deep Neural Network}
\newacronym{dr}{DR}{Dynamic Range}
\newacronym{fc}{FC}{Fully-Connected}
\newacronym{fcrit}{$P(f_{crit})$}{False Classification Probability}
\newacronym{fnoncrit}{$P(f_{non-crit})$}{Fault Propagation Probability}
\newacronym{ff}{FF}{Flip-Flop}
\newacronym{fi}{FI}{Fault Injection}
\newacronym{lut}{LUT}{Lookup Table}
\newacronym{mac}{MAC}{Multiply Accumulate}
\newacronym{nlf}{NLF}{Non-Linear Function}
\newacronym{rtl}{RTL}{Register Transfer Level}
\newacronym{sa}{SA}{Systolic Array}
\newacronym{sadnnacc}{SA-DNN accelerator}{Systolic Array based Deep Neural Network accelerator}
\newacronym{tv}{$t_v$}{Temporal Vulnerability} 

\glsdisablehyper
\makeglossaries

\title{Analysis of Single Event Induced Bit Faults in a Deep Neural Network Accelerator Pipeline}

\author[a,1]{Naïn Jonckers,\note{Corresponding author.}}
\author[a,b]{Toon Vinck,}
\author[a]{Peter Karsmakers}
\author[a]{and Jeffrey Prinzie}

\affiliation[a]{KU Leuven,\\Kleinhoefstraat 4, Geel, Belgium}
\affiliation[b]{Magics Technologies,\\Cipalstraat 3, Geel, Belgium}

\emailAdd{nain.jonckers@kuleuven.be}

\def\MNIST{MNIST}
\def\CIFARTEN{CIFAR-10}
\newcommand\bit[1]{#1-bit}
\newcommand\SAsize[1]{#1$\times$#1}

\abstract{%
    In recent years, the increased interest and the growth in application domains of \gls{ai}, and more specifically \glspl{dnn}, has led to an extensive usage of domain specific \gls{dnn} accelerator processors to improve the computational efficiency of \gls{dnn} inference. However, like any digital circuit, these processors are prone to faults induced by radiation particles such as heavy ions, protons, etc., making their use in harsh radiation environments a challenge.
    \par
    This work presents an in-depth analysis of the impact of such faults on the computational pipeline of a \gls{sadnnacc} by means of a \gls{rtl} \gls{fi} simulation in order to improve the observability of each hardware block. From this analysis, we present the sensitivity to single bit faults of register groups in the pipeline for three different \gls{dnn} workloads utilising two datasets, namely \MNIST{} and \CIFARTEN. These sensitivity figures are presented in terms of \gls{fnoncrit} and \gls{fcrit} which respectively show the probability that an injected fault causes a non-critical error (numerical offset) or a critical error (classification fault). From these results, we devise a fault mitigation strategy to harden the \gls{sadnnacc} in an efficient way, both in terms of area and power overhead.
}


\proceeding{Topical Workshop on Electronics for Particle Physics - TWEPP2025\\
  6-10 October 2025\\
  Rethymno, Crete, Greece}

\begin{document}
\notoc
\maketitle
\flushbottom
\clearpage
\glsresetall

\section{Introduction}
\label{sec:introduction}

Recent advancements in \gls{ai}, and specifically \glspl{dnn}, have led to machine implementations that often outperform human cognitive abilities. This has caused an increased interest in the usage of \glspl{dnn} in applications such as space satellite imagery \cite{bib:deep_learning_in_space} for, among others, bandwidth reduction and also particle physics \cite{bib:deep_learning_in_lhc}. These applications have challenging requirements due to bit faults induced by radiation and require either a form of physical hardening or circuit based hardening \cite{bib:deep_learning_in_space}.

Additionally, \gls{dnn} inference has high computational demands, requiring domain specific processors to compute these workloads with a good power efficiency, which is especially important in edge applications \cite{bib:survey_edge_accelerators}. However, previous work has shown that commercial \gls{dnn} accelerators such as Intel's Movidius Myriad X VPU and Google's Edge TPU are only suited for short duration low earth orbit missions \cite{bib:irradiation_movidius_tpu}, making their usage in high energy physics applications not feasible. To improve the reliability of such accelerators, techniques such as software-based mitigation or multiple redundant hardware processors are typically used \cite{bib:survey_dnn_reliability}, however, these techniques negatively impact the overall power efficiency \cite{bib:soft_errors_in_dnn_accelerators}. More efficient techniques usually focus on (1) efficient hardware mitigation using techniques such as razor flip-flops, error correcting codes, etc. or (2) fault-aware training of \glspl{dnn} \cite{bib:survey_dnn_reliability}. The latter approach leverages the inherent redundancy that neural networks can have when trained properly. However, due to the complex nature of \gls{dnn} models, propagation of faults on a hardware or \gls{dnn} accelerator level is often not well understood \cite{bib:error_propagation_in_dnn_accelerators,bib:reliablity_of_anns_for_space}.

To fill this knowledge gap, this work will perform a case study on a, often used, \bit{8} quantised \gls{sadnnacc} to analyse the fault propagation behaviour of \gls{dnn} workloads on a hardware level. This analysis can be achieved by either performing physical beam tests or by utilising a \gls{fi} simulation. While beam tests offer a more accurate representation of real-world conditions \cite{bib:reliablity_of_anns_for_space}, they provide limited insights into fault propagation within the accelerator logic due to physically limited on-chip signal observation. Although essential for evaluating reliability and fault tolerance, beam testing alone may not yield sufficient information for strategic fault mitigation. Therefore, in this work, we conduct a \gls{fi} injection simulation using a digital \gls{rtl} simulation environment utilising standard SystemVerilog \texttt{force} and \texttt{release} statements to inject faults on critical \gls{ff} nodes in the accelerator pipeline, allowing for a much better observability of fault propagation.

This paper is organized as follows: Section \ref{sec:architecture} describes the hardware architecture of the \gls{sadnnacc} and explains the data flow through this pipelined architecture. In section \ref{sec:test_methodology}, we present our \gls{fi} methodology and simulation. The results are presented and discussed in section \ref{sec:results} and finally, a conclusion together with potential future work is presented in section \ref{sec:conclusions}.

\section{Accelerator architecture}
\label{sec:architecture}

This work evaluates a \gls{sadnnacc} which includes, as the name implies, a \gls{sa}. \glspl{sa} were already described in 1979 \cite{bib:why_systolic_architecture}, but have only recently been widely used in \gls{dnn} accelerators such as the Google TPU \cite{bib:systolic_array_functional_safety}, due to their efficient computation of matrix and tensor products \cite{bib:survey_sa_based_accelerators}, which are an essential part of \glspl{dnn}. The employed \gls{sa} in this work's \gls{sadnnacc} is shown at the top of Figure \ref{fig:accelerator_pipeline_diagram} for a \SAsize{2} parametrisation (2 rows and 2 columns).

\begin{wrapfigure}[31]{r}{0.45\textwidth}
    \hspace*{-0.17\textwidth}
    \resizebox{0.68\textwidth}{!}{\input{fig/accelerator_pipeline.tex}}
    \caption{Block diagram of the \gls{sadnnacc} pipeline for a \SAsize{2} parametrisation}
    \label{fig:accelerator_pipeline_diagram}
\end{wrapfigure}

\noindent
Aside from matrix multiplications, which are used heavily in \gls{dnn} \gls{fc} or ``dense'' layers, \gls{conv2d} layers can also be computed by the \gls{sa}. This is accomplished by converting \gls{conv2d} operations into a matrix multiplication by means of convolution unrolling \cite{bib:conv_lowering}.

When computing these layers, activation inputs and layer weights are first loaded into the \gls{sa}. Since the employed \gls{sa} is weight-stationary, weights are kept statically for each (partial) matrix multiplication whilst activations are slanted and flow from left to right through \gls{mac} units. These perform one multiply accumulate calculation of the form $u_{r\times c}^{i32} = a_{r}^{i8} w_{r\times c}^{i8}$ into a \bit{32} register to account for overflow, with $a_r^{i8}$ a signed \bit{8} activation input for row $r$, $w_{r\times c}^{i8}$ a stationary signed \bit{8} weight for row $r$ and column $c$ and $u_{r\times c}^{i32}$ a signed \bit{32} partial result for row $r$ and column $c$. Thus partial results flow from the top to the bottom of the \gls{sa}. Often the parametrisation of the \gls{sa} (\SAsize{2} in Figure \ref{fig:accelerator_pipeline_diagram}) will be smaller than the actual matrix multiplication. To handle this, a column-wide \bit{32} accumulator is foreseen which can also add an additional bias $b_{c}^{i32}$ to the final layer result. Thus, for each column $c$, Equation \ref{eqn:mmult} is computed, where $R$ is the total number of rows, resulting in a total vector-matrix multiplication over each column $c$:
\begingroup
\vspace*{-0.3em}
\begin{equation}\label{eqn:mmult}
    {y'}_{c}^{i32} = b_{c}^{i32} + \sum_{r=0}^{R-1} \left( a_{r}^{i8} w_{r\times c}^{i8} \right)
\end{equation}
\endgroup

\noindent
After accumulating the \bit{32} partial sums from the \gls{sa} in the accumulators for each column, the result is rescaled to an \bit{8} resolution in the ``rounding'' block. This is required since all memories in the \gls{sadnnacc} are \bit{8} quantised (except for the bias) as mentioned before. Hence the final result should also be rescaled to this resolution.

To accomplish this, the rounding block applies a scale factor $s$ to the \bit{32} accumulator result. This scale factor is constrained to a power of two so that the usual division by this scale factor, becomes a simple right shift $S = \log_2(1/s)$ in hardware \cite{bib:quant_training_dnn,bib:dnn_quantization_whitepaper}. Furthermore, the shifted result is rounded and clipped to the \bit{8} signed integer range. This operation can thus contribute positively to fault masking by discarding least significant bits of the \bit{32} accumulated result (shift operation) as well as discarding some most significant bits of this result (clipping). This leads to Equation \ref{eqn:rounding} where the matrix multiplication result is re-quantised to \bit{8} using this rounding operation.
\begin{equation}\label{eqn:rounding}
    y_{c}^{i8} = \text{clip} \left(
        \text{round} \left(
            b_{c}^{i32} + \sum_{r=0}^{R-1} \left( a_{r}^{i8} w_{r\times c}^{i8} \right)
        \right)
    \right) >> S
\end{equation}

Finally, after the rounding stage, a \gls{nlf} block is implemented which can compute any non-linear function by utilizing a \gls{lut} structure. After this \gls{nlf} stage, a final pooling stage is foreseen which can compute max-pool layers, commonly employed by many \glspl{dnn}, and specifically \glspl{cnn}. This unit can be bypassed since pooling is not required after each \gls{dnn} layer.

\section{Test methodology}
\label{sec:test_methodology}
\vspace{-0.2em}

\begin{wrapfigure}{r}{0.55\textwidth}
    \vspace{-1.2em}
    \hspace{-2.5em}
    \resizebox{0.66\textwidth}{!}{
{
\renewcommand*{\ttdefault}{qcr}
\fontfamily{qhv}\selectfont

\def\styLineSize{0.05cm}
\pgfmathsetmacro\styLineSizeHalf{\styLineSize / 2}
\tikzstyle{sty-arrow}=[line width=0.1cm, >=stealth]
\tikzstyle{sty-pipeline}=[draw, line width=\styLineSize, fill=green!20!yellow!20!white]
\tikzstyle{sty-ecc-mem}=[draw, line width=\styLineSize, densely dashdotted, fill=red!40!white]
\tikzstyle{sty-tmr-logic}=[draw, line width=\styLineSize, densely dashed, fill=red!60!blue!20!white]
\tikzstyle{sty-iopad}=[draw, squarecross, line width=\styLineSize,
                       minimum width=0.5cm, minimum height=0.5cm, inner sep=0pt]

\begin{tikzpicture}
    \node[draw, dashed, minimum width=5.5cm, minimum height=11.5cm, anchor=north west]
          (pipeline-detail) at (0,0) {};
    \node[anchor=north west] (pipeline-detail-text) at (pipeline-detail.north west)
         {\large Data pipeline};

    \node[sty-pipeline, minimum width=3.5cm, minimum height=3.5cm, anchor=north,
          text width=3.5cm, align=center, inner sep=0pt, yshift=-1.5cm, xshift=0.5cm]
          (pipeline-detail-sa) at (pipeline-detail.north) {Systolic Array};
    \node[sty-pipeline, minimum width=3.5cm, minimum height=0.8cm, anchor=north,
          text width=3.5cm, align=center, inner sep=0pt, yshift=-0.5cm]
          (pipeline-detail-acc) at (pipeline-detail-sa.south) {Accumulator};
    \node[sty-pipeline, minimum width=3.5cm, minimum height=0.8cm, anchor=north,
          text width=3.5cm, align=center, inner sep=0pt, yshift=-0.5cm]
          (pipeline-detail-round) at (pipeline-detail-acc.south) {Rounding};
    \node[sty-pipeline, minimum width=3.5cm, minimum height=0.8cm, anchor=north,
          text width=3.5cm, align=center, inner sep=0pt, yshift=-0.5cm]
          (pipeline-detail-nlf) at (pipeline-detail-round.south) {Non-Linear Function};
    \node[sty-pipeline, minimum width=3.5cm, minimum height=0.8cm, anchor=north,
          text width=3.5cm, align=center, inner sep=0pt, yshift=-0.5cm]
          (pipeline-detail-pool) at (pipeline-detail-nlf.south) {(Optional) pooling};

    \node[sty-ecc-mem, minimum width=1cm, minimum height=0.5cm, rotate=90, yshift=0.7cm]
          (pipeline-detail-amem) at (pipeline-detail-sa.west) {AMEM};
    \node[sty-ecc-mem, minimum width=1cm, minimum height=0.5cm, rotate=90, yshift=0.7cm]
          (pipeline-detail-bmem) at (pipeline-detail-acc.west) {BMEM};
    \node[sty-ecc-mem, minimum width=1cm, minimum height=0.5cm, yshift=0.7cm]
          (pipeline-detail-wmem) at (pipeline-detail-sa.north) {WMEM};
    \node[sty-ecc-mem, minimum width=1.5cm, minimum height=0.5cm, yshift=-0.7cm]
          (pipeline-detail-amemwb) at (pipeline-detail-pool.south) {AMEM (write back)};

    \draw[sty-arrow, ->] (pipeline-detail-wmem) -- (pipeline-detail-sa.north);
    \draw[sty-arrow, ->] (pipeline-detail-amem) -- (pipeline-detail-sa.west);
    \draw[sty-arrow, ->] (pipeline-detail-bmem) -- (pipeline-detail-acc.west);
    \draw[sty-arrow, ->] (pipeline-detail-pool.south) -- (pipeline-detail-amemwb);

    \draw[sty-arrow, ->] (pipeline-detail-sa.south) -- (pipeline-detail-acc.north);
    \draw[sty-arrow, ->] (pipeline-detail-acc.south) -- (pipeline-detail-round.north);
    \draw[sty-arrow, ->] (pipeline-detail-round.south) -- (pipeline-detail-nlf.north);
    \draw[sty-arrow, ->] (pipeline-detail-nlf.south) -- (pipeline-detail-pool.north);

    \node[draw, dashed, minimum width=19cm, minimum height=16.5cm, anchor=north west]
          (framework) at ([shift={(-1,1)}]pipeline-detail.north west) {};
    \node[anchor=north] (framework-text) at (framework.north) {\LARGE Fault injection framework};
    \begin{scope}[on background layer]
    \node[draw, thick, anchor=north west, xshift=0.5cm, minimum width=11cm, text width=11cm,
    inner sep=8pt, align=left, fill=white] (framework-code) at (pipeline-detail.north east) {%
        \ttfamily
        RTL.WMEM = dnn\_model.weights;
        \\
        RTL.BMEM = dnn\_model.biases;
        \\\ \\
        \textbf{for} image \textbf{in} image\_set
        \\
        \quad RTL.AMEM = image;
        \\
        \quad \textcolor{brown}{// Obtain golden reference}
        \\
        \quad \textbf{for} cycle \textbf{in} model\_cycles
        \\
        \quad\quad \#T\textsubscript{clk/2} clk = 0;
        \\
        \quad\quad \#T\textsubscript{clk/2} clk = 1;
        \\
        \quad golden = RTL.amem\_wb;
        \\
        \quad \textcolor{brown}{// log golden data}
        \\
        \quad \textbf{log}(image, model\_cycles, golden);
        \\\ \\
        \quad \textcolor{brown}{// Fault injection}
        \\
        \quad \textbf{for} iter \textbf{in} iterations
        \\
        \quad\quad fault\_ff = \$random(RTL.ff\_nodes);
        \\
        \quad\quad fault\_cycle = \$random(0, dnn\_model.cycles);
        \\
        \quad\quad \textbf{for} cycle \textbf{in} (0,...,dnn\_model.cycles)
        \\
        \quad\quad\quad \textbf{if} cycle == fault\_cycle
        \\
        \quad\quad\quad\quad fault\_inject(RTL.node(fault\_ff));
        \\
        \quad\quad\quad \#T\textsubscript{clk/2} clk = 0;
        \\
        \quad\quad\quad \#T\textsubscript{clk/2} clk = 1;
        \\
        \quad\quad injected = RTL.amem\_wb;
        \\
        \quad\quad \textcolor{brown}{// log fault injection data}
        \\
        \quad\quad \textbf{log}(iter, fault\_cycle, fault\_ff, injected);
    };

    \draw[sty-arrow, ->, color=cyan]
         ([shift={(0.2,-0.4)}]framework-code.north west)
         to[out=180,in=90] (pipeline-detail-wmem.north);
    \draw[sty-arrow, ->, color=cyan]
         ([shift={(0.2,-1)}]framework-code.north west) to[out=200,in=0] +(-5,-0.3)
         to[out=180,in=180] (pipeline-detail-bmem.north);
    \draw[sty-arrow, ->, color=cyan]
         ([shift={(0.7,-2.55)}]framework-code.north west)
         to[out=180,in=180] (pipeline-detail-amem.north);
    \draw[sty-arrow, ->, color=green!40!blue]
         (pipeline-detail-amemwb.south)
         to[out=-70,in=-60] ([shift={(13.5,0)}]pipeline-detail-amemwb.south)
         to[out=120,in=0] ([shift={(6.4,1.5)}]framework-code.south west);
    \draw[sty-arrow, ->, color=green!40!blue]
         ([shift={(13.5,0)}]pipeline-detail-amemwb.south)
         to[out=60,in=0] ([shift={(5.4,8.35)}]framework-code.south west);

    \foreach \pipelinenode in {pipeline-detail-sa,pipeline-detail-acc,pipeline-detail-round,pipeline-detail-nlf,pipeline-detail-pool} {
        \draw[sty-arrow, ->, color=red]
             ([shift={(1.7,3.1)}]framework-code.south west)
             to[out=180,in=0] (\pipelinenode.east);
    }

    \end{scope}
\end{tikzpicture}
}}
    \vspace{-1.2cm}
    \caption{Simulation \gls{fi} framework}
    \label{fig:simulation_fi_setup}
    \vspace{-0.2cm}
\end{wrapfigure}
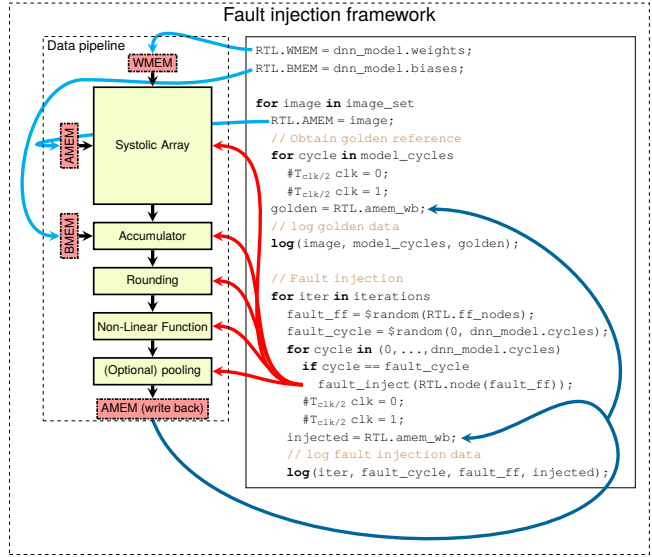

In order to improve the observability of fault propagation in data pipeline registers, this work performs several \gls{fi} experiments using \gls{rtl} simulations. The constructed framework for these experiments is highlighted in Figure \ref{fig:simulation_fi_setup}. The right-hand side of this figure shows a pseudo-algorithm of the \gls{fi} framework and the left-hand side shows the simplified accelerator data pipeline from Figure \ref{fig:accelerator_pipeline_diagram}, where AMEM, WMEM and BMEM respectively hold the model's input/output activations ($a_{r}^{i8}$, $y_{c}^{i8}$), weights ($w_{r\times c}^{i8}$) and biases ($b_{c}^{i32}$).

For each selected \gls{dnn} model, we first load the model's weights and biases into their respective memories. Then, for several input images we perform multiple inferences, where each first inference computes the golden reference (i.e. no fault injected) for that specific input image. Consecutive inferences for the same image will then inject a fault at (1) a random cycle and (2) a random register \gls{ff}. The \gls{dnn} output after each of these \gls{fi} inferences is then compared to the golden reference to assess fault propagation for that specific register \gls{ff}. The following  \glspl{dnn} were used based on a trade-off between the time needed for statistically relevant \gls{fi} simulations and representativeness to typical \gls{dnn} and \gls{dnn} layer usage:

\vspace{-0.7em}
\begin{itemize}
    \item \textbf{3L-\acrshort{fc} + \MNIST}: 3-Layer \gls{fc} network with intermediate ReLU activation function, trained on \MNIST{} (post-quantisation accuracy: 96.9\%)
    \vspace{-0.8em}
    \item \textbf{LeNET + \MNIST}: LeNET with ReLU and max-pooling, trained on \MNIST{} (post-quantisation accuracy: 98.6\%)
    \vspace{-0.8em}
    \item \textbf{LeNET + \CIFARTEN}: Modified (increased to 3 input channels) LeNET with ReLU and max-pooling, trained on \CIFARTEN{} (post-quantisation accuracy: 70.8\%)
\end{itemize}
\vspace{-0.7em}

\noindent
When comparing the classification result of an inference to the golden reference, three situations can occur: (i) No difference is observed due to the model's redundancy or fault masking by the post-processing stages such as rounding and \gls{nlf} activation (e.g. ReLU). (ii) An injected fault may cause a numerical difference on the \gls{dnn} model's output logits but not cause any classification corruption. In this case, we classify the fault in the \gls{fnoncrit} category, which is the likelihood that an injected fault for a specific register, causes a non-critical \gls{dnn} output fault. (iii) An injected fault causes a classification corruption in which case we classify it in the \gls{fcrit} category, which is the likelihood that an injected register fault causes a critical \gls{dnn} output fault.

\section{Results}
\label{sec:results}

\definecolor{colFlowAct}{HTML}{0000FF}
\definecolor{colFlowPart}{HTML}{FF0000}
\definecolor{colFFchainH}{HTML}{6666DD}
\definecolor{colSAregH}{HTML}{CCCCFF}
\definecolor{colWeights}{HTML}{EEAA44}
\definecolor{colSAregV}{HTML}{FFCCCC}
\definecolor{colFFchainV}{HTML}{DD6666}
\definecolor{colAccum}{HTML}{AA0000}
\definecolor{colRound}{HTML}{99CC77}
\definecolor{colNLF}{HTML}{779977}
\definecolor{colPool}{HTML}{997799}

\begin{figure*}
    \vspace{-1em}
    \hspace*{-0.05\textwidth} 
    \begin{subfigure}{0.6\textwidth}
        (a) \raisebox{-0.6\height}{\includegraphics[width=0.9\textwidth]{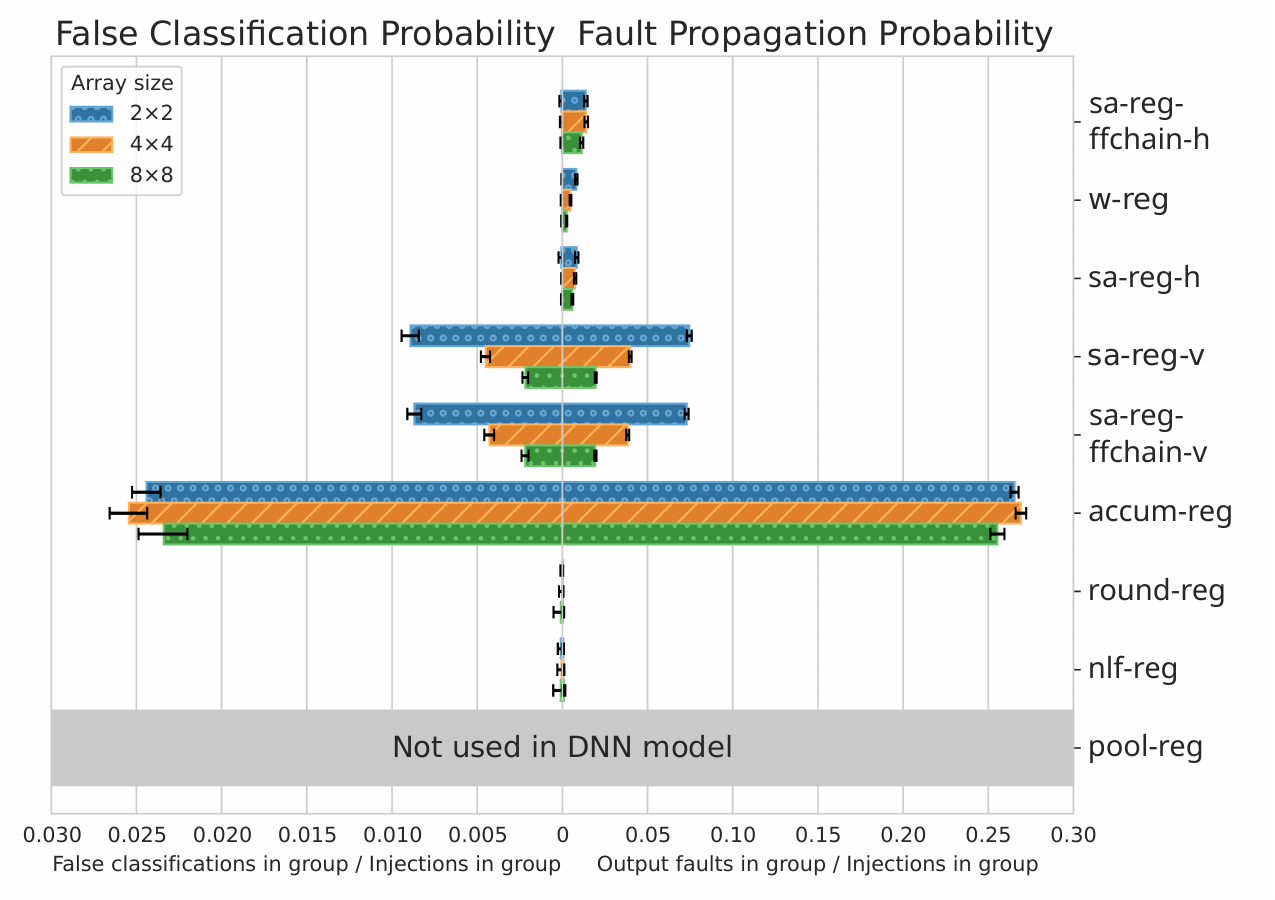}}
    \end{subfigure} \begin{subfigure}{0.6\textwidth}
        (b) \raisebox{-0.6\height}{\includegraphics[width=0.9\textwidth]{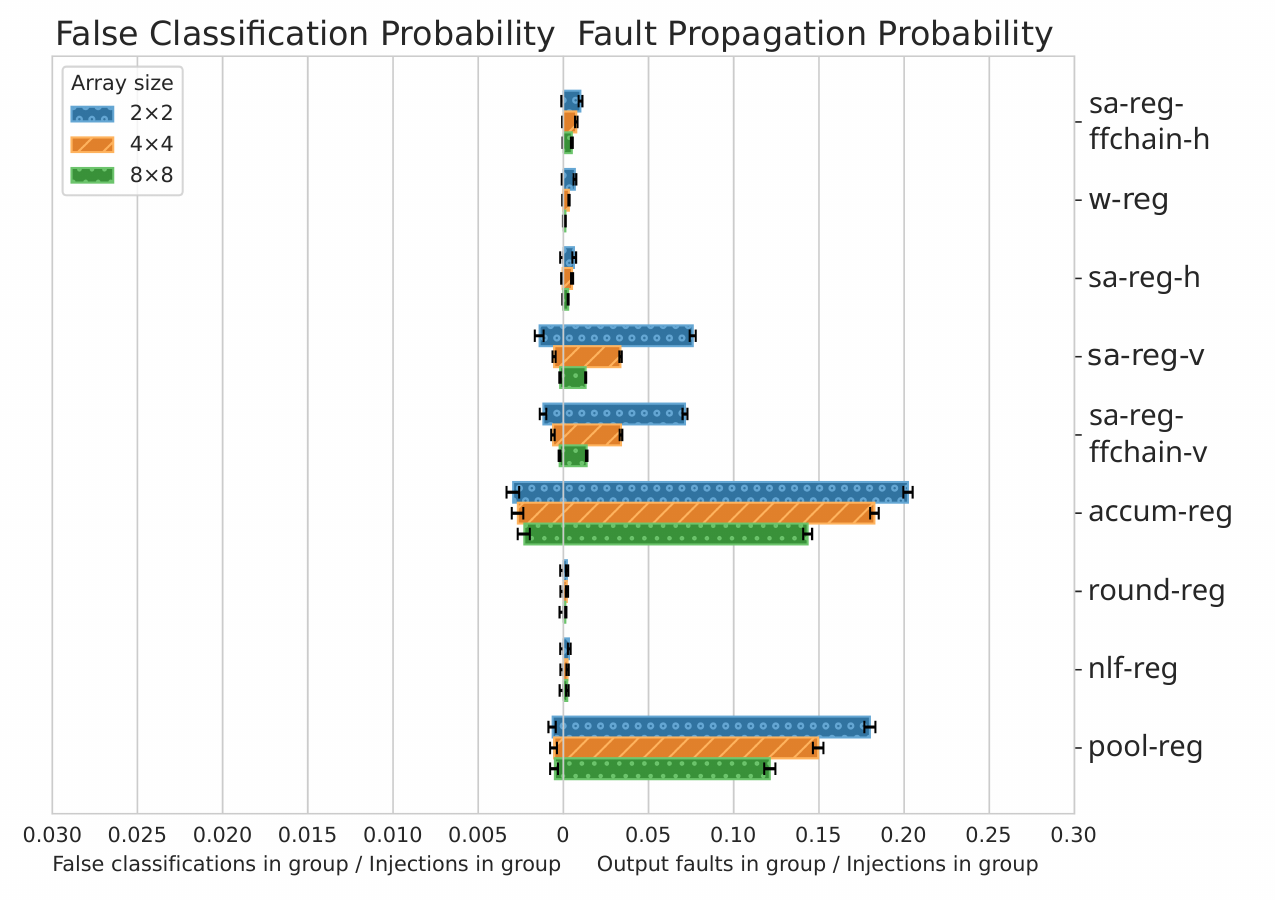}}
    \end{subfigure}
    \hspace*{0.2\textwidth} 
    \begin{subfigure}{0.6\textwidth}
        (c) \raisebox{-0.6\height}{\includegraphics[width=\textwidth]{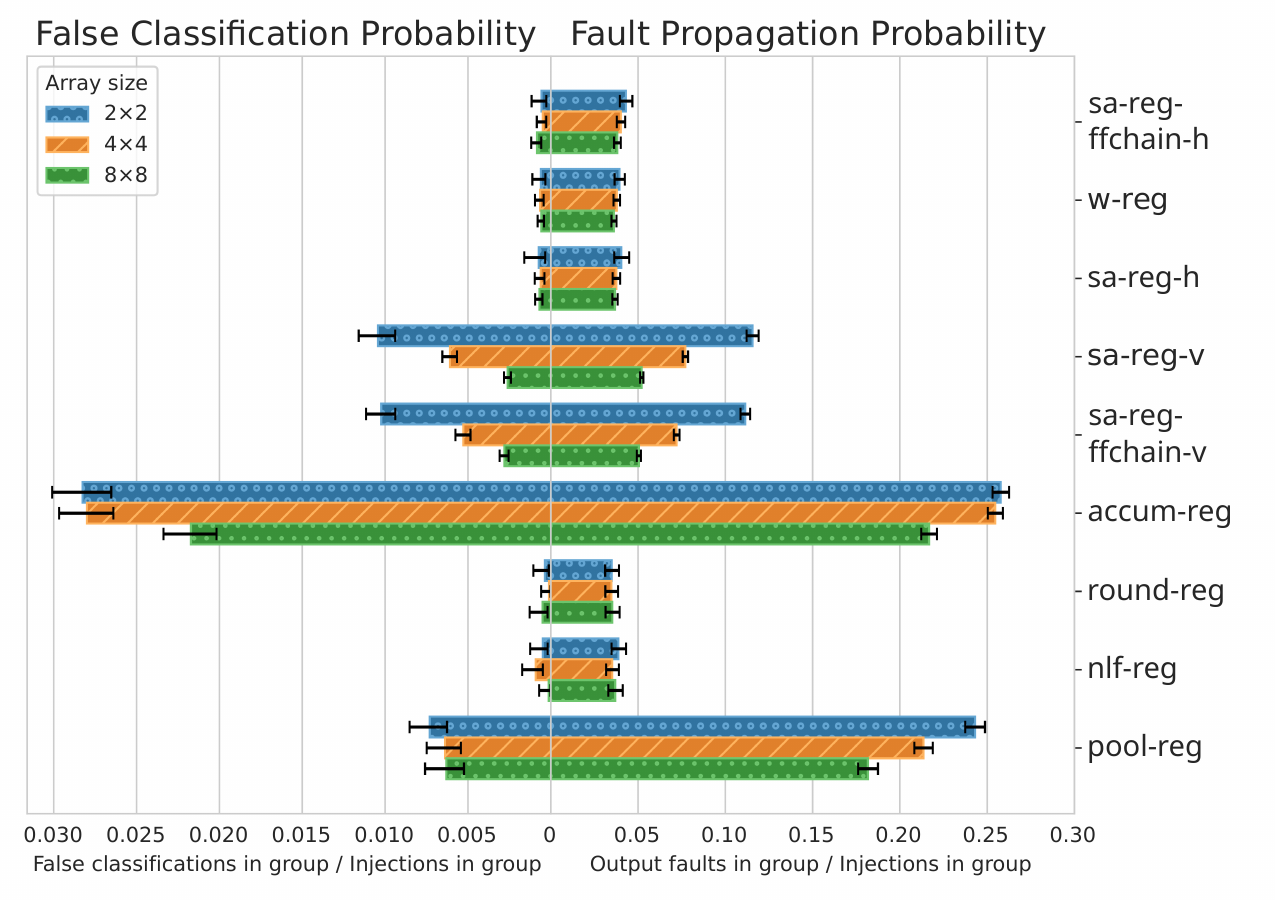}}
    \end{subfigure}
    \vspace{-1em}
    \caption{\acrshort{dnn} test results for: (a) 3-Layer \acrshort{fc} network with \MNIST, (b) LeNET with \MNIST{} and (c) LeNET with \CIFARTEN}
    \label{fig:results_dnn}
    \vspace{-1em}
\end{figure*}

Figure \ref{fig:results_dnn} shows the \gls{fi} results for the three \glspl{dnn}. The right-hand side of each figure shows \gls{fnoncrit} for each register whereas the left-hand side shows \gls{fcrit}. Across all \glspl{dnn}, the impact of faults appears to be similar across four groups of registers. To simplify the discussion of the results, we propose four groups: (i) \textit{\bit{8} \gls{sa} registers} (\texttt{\color{colWeights!80!red}w-reg}, \texttt{\color{colFFchainH!80!blue}sa-ffchain-h-reg}, \texttt{\color{colSAregH!80!blue}sa-h-reg}), (ii) \textit{\bit{32} \gls{sa} registers} (\texttt{\color{colFFchainV!70!red}sa-ffchain-v-reg}, \texttt{\color{colSAregV!70!red}sa-v-reg}), (iii) \textit{\bit{32} accumulator registers} (\texttt{\color{colAccum!95!black}accum-reg}) and (iv) \textit{post-processing registers} (\texttt{\color{colRound!80!green}round-reg}, \texttt{\color{colNLF!80!black}nlf-reg}, \texttt{\color{colPool!80!violet}pool-reg}).

For the first \textbf{3L-\acrshort{fc} network, trained on \MNIST}, \gls{dnn} in Figure \ref{fig:results_dnn} (a), the first observation is that neither the \bit{8} \gls{sa} nor the post-processing registers contribute significantly to \gls{fnoncrit} and \gls{fcrit} (note that pooling is not used in this model). This is due to their small (\bit{8}) \gls{dr} and the fact that the post-processing registers can mask faults by shifting (\texttt{round-reg}) and \gls{nlf} activation (such as ReLU). In contrast, the \bit{32} \gls{sa} registers have much higher fault probabilities up to about 25\% \gls{fcrit} for the \bit{32} accumulator. This can be attributed to their high \gls{dr}, which increases the probability that a single-bit fault, during inference, causes an output difference large enough to cause a \gls{dnn} misclassification. The accumulator register has an even higher \gls{fnoncrit} and \gls{fcrit}, which can be explained by the fact that it has a large \gls{tv} when accumulating many partial matrix multiplication results over a longer period of time. Finally, it is clear that the \bit{32} \gls{sa} register sensitivity decreases as the \gls{sa} size is scaled up. Fortunately, a larger array is also more efficient for calculating large matrix multiplications which are prevalent in modern \gls{dnn} models.

When looking at \textbf{LeNET with \MNIST}, a \gls{cnn} model with pooling layers, in Figure \ref{fig:results_dnn} (b), we observe a large decrease in \gls{fcrit} from around 25\% in the 3L-\gls{fc} network to a maximum of about 2.5\% in this network even though the values for \gls{fnoncrit} are in the same order of magnitude, albeit slightly lower. This can be attributed to an increased model redundancy of LeNET compared to a simple 3L-\gls{fc} network. Furthermore, we now also observe a large \gls{fnoncrit} of almost 18\% for a \SAsize{2} \gls{sa} in the pooling registers. This can be attributed to a larger \gls{tv} of this register since it is also a form of accumulator register. However due to its small \gls{dr} of \bit{8} and the fact that pooling layers are used only as the first LeNET layers, the impact on \gls{fcrit} is again negligible like the rest of the post-processing registers.

Finally, \textbf{LeNET with \CIFARTEN}, highlighted in Figure \ref{fig:results_dnn} (c), exhibits increased fault probabilities compared to LeNET with MNIST, where the worst-case \gls{fcrit} reaches up to 30\% for the accumulator register. Rather than being the cause of the switch to a different dataset (\CIFARTEN), it is more probable that this increased \gls{fcrit} is due to a reduced training accuracy compared to LeNET with \MNIST{} (70.8\% v.s. 98.6\%). \CIFARTEN{} exhibits a more complex classification task and thus increases the uncertainty of the model, leading to al lower accuracy and a reduced margin for error \cite{bib:dnn_strategic_input_selection}. As such, smaller numerical errors are more likely lead to misclassifications. This highlights that the training process also strongly impacts fault propagation during \gls{dnn} inference.

\section{Conclusions}
\label{sec:conclusions}

In this work, we performed \gls{rtl} \gls{fi} simulations on a \gls{sadnnacc} to increase observability of fault propagation in internal pipeline registers. We observed that, on a \textbf{hardware level}, the main contributor to an increased fault propagation probability, both in terms of \gls{fnoncrit} and \gls{fcrit}, was (1) the \gls{dr} and (2) \gls{tv} of a register. Hence, the \bit{32} \gls{sa} and the \bit{32} accumulator registers are the most vulnerable. Furthermore, on a \textbf{model level}, it was observed that from the 3L-\gls{fc} model to LeNET (both trained on \MNIST{}) a factor 10 decrease in \gls{fcrit} occurred, highlighting that the \gls{dnn} model's inherent redundancy also plays an important role in fault propagation. Finally between LeNET trained on \MNIST{} and LeNET trained on \CIFARTEN{}, an increase of more than 10 times is observed due to the latter model's lower training accuracy (70.8\% v.s. 98.6\%), highlighting that the model's training phase is also very important to the overall fault propagation.

From these results it is clear that, in order to obtain a power and area efficient fault tolerant \gls{dnn} accelerator, one needs to take both hardware and the software \gls{dnn} model into account. By leveraging proper \gls{dnn} model training fault propagation can already be reduced significantly in software. Further faults can efficiently be detected in hardware using for instance a low-overhead fault check mechanism \cite{bib:dnn_accelerator_check_neuron} and then optionally recomputed. By leveraging this hardware-software co-design a very efficient, fault tolerant \gls{ai} accelerator can be designed.

\clearpage
\bibliography{references}

\end{document}